\documentclass[submission,copyright,creativecommons]{eptcs}
\providecommand{\event}{ICE 2016} 
\usepackage{breakurl}             

\usepackage{amssymb}
\usepackage{graphicx}

\usepackage{color}
\usepackage{mathpartir}
\usepackage{amsmath}
\usepackage{graphicx}
\usepackage{algorithm}
\usepackage[noend]{algpseudocode}
\usepackage{listings}
\usepackage{url}
\newtheorem{theorem}{Theorem}
\urldef{\mailsa}\path|{k.azadbakht, f.s.de.boer, v.n.serbanescu}@cwi.nl| 

\title{Multi-Threaded Actors}
\author{Keyvan Azadbakht \qquad\qquad Frank S. de Boer \qquad\qquad Vlad Serbanescu
\institute{Centrum Wiskunde and Informatica\\
Amsterdam, Netherlands}
\email{\{k.azadbakht, f.s.de.boer, v.n.serbanescu\}@cwi.nl}
}
\def\titlerunning{Multi-Threaded Actors}
\def\authorrunning{K. Azadbakht, F. S. de Boer \& V. Serbanescu}
\begin{document}
\maketitle
\begin{abstract}
In this paper we introduce a new programming model of multi-threaded actors which feature the parallel processing of their messages. In this model an actor consists of a group of active objects which share a message queue. We provide a formal operational semantics, and a description of a Java-based implementation for the basic programming abstractions describing multi-threaded  actors. Finally, we evaluate our proposal by means of an example application.
\end{abstract}

\section{Introduction}

Object-oriented programs organize data and corresponding operations by means of a hierarchical structure of classes.
A class can be dynamically instantiated and as such extends the concept of a module.
Operations are performed by corresponding method calls on class instances, namely objects.
In most object-oriented languages, like Java, method calls are executed by a thread of control which gives rise
to a stack of call frames. In a distributed setting, where objects are instantiated over different machines, remote method calls
involve a \emph{synchronous rendez-vous} between caller and callee.

It is generally recognized that \emph{asynchronous} communication is better suited for distributed applications.
In the Actor-based programming model of concurrency \cite{Agha90} actors communicate via asynchronous messages.
In an object-oriented setting
such a message specifies a method of the callee and includes the corresponding actual parameters.
Messages in general are queued
and trigger execution of the body of the specified method by the callee, when dequeued.
The caller object proceeds with its own execution and may synchronize on the return value by means of \emph{futures} \cite{BoerCJ07}.

In \cite{SchaferP10} JCoBox, a Java extension with an actor-like concurrency model based on the notion of concurrently running object groups, the concept of coboxes is introduced which integrates thread-based synchronous method calls
with asynchronous communication of messages in a \emph{Global Asynchronous, Local Synchronous (GALS)} manner.
More specifically, synchronous communication of method calls is restricted to objects belonging to the same cobox.
Objects belonging to the same cobox share control, consequently within a cobox at most one thread of synchronous method calls is executing.
Only objects belonging to different coboxes can communicate via asynchronous messages.

Instead of sharing control, 
in this paper we introduce an Actor-based language which features new programming abstractions
for  parallel processing of  messages. The basic distinction the language supports is that between the instantiation of an Actor  class which gives rise to 
the initialization of a group of active objects  sharing a queue and that which adds a new active object to an existing group.
Such a group of active objects sharing a message queue constitutes a multi-threaded actor which features
the parallel processing of its messages.
The  distinction between actors and active objects is reflected by the type system which includes an explicit type for actors and which is used to restrict
the  communication between actors to asynchronous method calls.
In contrast to the concept of a cobox, a group of active objects sharing a queue has its own distinct identity (which coincides with the initial active object).
This distinction further allows, by means of simple typing rules,
to restrict the communication between active objects to synchronous method calls.
When an active object  fetches a message from the shared message queue, the object starts executing a corresponding thread in parallel
with all the other threads.
This basic mechanism gives rise to the new
programming concept of a \emph{Multi-threaded Actor (MAC)} which provides
a powerful Actor-based abstraction of the notion of a \emph{thread pool}, as for example, implemented by the Java library \emph{java.util.concurrent.ExecutorService}.
We further extend the concept of a MAC with a powerful high-level concept of synchronized data to
constrain the parallel execution of messages.

In this paper we provide a formal operational semantics like Plotkin \cite{Plotkin04a}, and a description of a Java-based implementation
for the basic programming abstractions
describing sharing of message queues between active objects.
The proposed run-time system is based on the ExecutorService interface and the use of \emph{lambda expressions} in
the implementation of asynchronous execution and messaging.

\paragraph{Related work}

Since Agha introduced in \cite{Agha90} the basic Actor model of concurrent computation in distributed systems, a great variety of Actor-based programming languages have
been developed. In most of these languages, e.g., Scala \cite{Haller2009202}, Creol \cite{JohnsenO07}, 
ABS \cite{johnsen10fmco}, JCoBox \cite{SchaferP10}, Encore \cite{brandauer2015parallel}, 
ProActive \cite{Caromel08}, AmbientTalk \cite{MBDM07}, Rebeca \cite{Sirjani06}, actors execute messages stored in their own message queue.
The Akka library for Actor-based programming however does support sharing of
message queues between actors.
In this paper we introduce a new corresponding Actor-based programming abstraction
which integrates a thread-based execution of messages with event-based asynchronous message passing.

Our work complements in a natural manner that of \cite{SchaferP10} which
introduces groups of actors sharing control.
Another approach to extending the Actor-based concurrency model is that of Multi-threaded active objects (MAO) \cite{henrio2013multi} and Parallel
Actor Monitors (PAM) \cite{ScholliersTM14} which allow the parallel execution of the different method invocations within an actor. Another approach is followed in the language Encore which provides an explicit construct for describing parallelism within the execution of one method \cite{part2016}.
In contrast to these languages, we do allow the parallel execution of different asynchronous method invocation inside a group of active objects which provides an overall functionality as that of an actor, e.g., it supports an interface for asynchronous method calls and a unique identity. 
Further we provide a new high-level language construct for specifying that certain parameters of a method are \textit{synchronized}, which allows a fine-grained parameter-based scheduling of messages. In contrast, the more coarse-grained standard scheduling of methods as provided by Java, PAM, and MAO, and JAC \cite{haustein2006jac} in general only specify which methods can run in parallel independent of the actual parameters. 
 \cite{williams2001method} also shows the notion of Microsoft COM (Component Object Model)'s multi-threaded apartment. In this model, calls to methods of objects in the multi-threaded apartment can be run on any thread in the apartment. It however lacks the ability of setting scheduling strategies (e.g. partial order of incoming messages in the next section).
Multi-threaded actors offer  a higher level of abstraction to parallel programming and can be viewed as similar to the OpenMP \cite{dagum1998openmp} specification for parallel programming in C, C++ and Fortran. 

\par The rest of this paper is organized as follows: In section \ref{mot}, an application example is established, by which we introduce the key features of MAC. Section \ref{syn} describes the syntax of MAC and the type system. Section \ref{opr} presents the operational semantics.
In Section \ref{exp} we show the implementation of MAC in the Java language and explain its features through an example. We draw some conclusions in Section \ref{concl} where we briefly discuss extensions and variations describing static
group interfaces,  support for the cooperative scheduling of the method
invocations within an actor (as described in for example \cite{JohnsenO07}), synchronization between the threads of a MAC, and encapsulation of the active objects belonging to the same actor.

\section{Motivating Example}\label{mot}
In this section, we explain an example which is used in the rest of the paper to show the notion of MAC. We also raise a challenge regarding this example which is solved later in our proposed solution. We present a simple concurrent bank service where the requests such as withdrawal, checking, and transferring credit on bank accounts are supported. The requests can be submitted in parallel by several clients of the bank. The system should respect the temporal order of the submitted requests on the same accounts. For instance, checking the credit of an account should return the amount of credit for the account after withdrawal, if there is a withdrawal request for that account which precedes the check request. The requests can be sent asynchronously. Therefore, respecting temporal order of two events means that there is a happens-before relation between termination of the execution of the former event and starting the execution of the latter. 

Existing technologies are either not able to implement this property or they need ad-hoc explicit synchronization mechanism which can be complicated and erroneous. Using locks on accounts (e.g. \textit{synchronized} block in Java) may cause deadlock or violate the ordering, unless managed explicitly at the lower level, since two accounts are involved in transferring credit. Another approach is to implement the scheduler in PAM \cite{ScholliersTM14} to support such ordering which raises synchronization complexities. The last alternative we investigate in this section is that to implement the service as a thread pool (e.g. ExecutorService in Java), where the above ordering is respected explicitly via passing the \textit{future} variable corresponding to the previous task, to the current one. The variable is then used to force the happens-before relation by suspending the process until the future is resolved (e.g. \textit{get} method in Java). One challenge is that the approach requires that the submitter knows and has access to the future variables associated to the previous task (or tasks in case the task being submitted is a \textit{transfer}). The other challenge is that, in a parallel setting with multiple concurrent source of task submitters, how to provide such knowledge. Last but not least, the approach first activates the task by allocating a thread and then the task may be blocked which imposes overhead, while a desirable solution forces the ordering upon the task activation. As shown in the rest of the paper, we provide the notion of MAC which overcomes this issue only via annotating the parameters based on which we aim to respect the temporal order.

\section{Syntax of MAC}
\label{syn}
Figure \ref{syntax} specifies the syntax. 
A \emph{MAC} program \emph{P} defines interfaces and classes, and a main statement. 
An interface \emph{IF} has a name \emph{I} and method signatures \emph{Sg}. A class \emph{CL} has a name \emph{C}, interfaces $\overline{I}$ that \emph{C} implements (that specify the possible types for its instances), formal parameters and attributes $\overline{x}$ of type $\overline{T}$, and methods $\overline{M}$. 
A multi-threaded actor consisting of a group of active objects (a MAC)  which share a queue of messages of type $I$ is denoted by
\textbf{Actor}\texttt{<}I\texttt{>}.
The type \textbf{Fut}\texttt{<}T\texttt{>} denotes futures which store return values of type $T$.
The \emph{fields} of the class consist of both its parameters and its attributes. A method signature \emph{Sg} declares a method with name \emph{m} and the formal parameters $\overline{x}$ of types $\overline{T}$ with optional \textbf{sync}\texttt{<}l\texttt{>} modifier
which is used to indicate that the corresponding parameter contains synchronized data.   
The user-defined label l allows to introduce different locks  for the same data type. Informally, a message which consists of such synchronized data can only be activated 
if the specified data has not been locked.

\begin{figure}[h]
\begin{align*}
T::=\ &\ \text{Bool}\ |\ I\ |\ \textbf{Actor}\texttt{<}I\texttt{>}\ |\ \textbf{Fut}\texttt{<}T\texttt{>} \\ 
P::=\ &\overline{IF}\ \overline{CL}\ \{\overline{T}\ \overline{x}; s\}\\
CL::=\ &\textbf{class}\ C [(\overline{T}\ \overline{x})]\ \textbf{implements}\ \overline{I} \{\overline{T}\ \overline{x}; \overline{M}\}\\
IF::=\ &\textbf{interface}\ I \{\overline{Sg}\}\\
Sg::=\ &\overline{[\textbf{sync}\texttt{<}l\texttt{>}]}\ T\ m(\overline{[\textbf{sync}\texttt{<}l\texttt{>}]}\ \overline{T}\ \overline{x})\\
M::=\ &Sg\{\overline{T}\ \overline{x}; s\}\\
s::=\ &\ x=e\ |\ s;s\ |\ e.\textbf{get}\ |\ \textbf{if}\ b\{s_1\}\textbf{else}\{s_2\}\ |\ \textbf{while}\ b\{s\}\\
e::=\ &\textbf{null}\ |\ b\ |	\ x\ |\ \textbf{this}\ |\ \textbf{new }[\textbf{actor}]\ C[(\overline{e})]\ |\ e.m(\overline{e})\ |\ e!m(\overline{e})\\
b::=\ &e?\ |\ b\ |\ b\land b\\
\end{align*}
\caption{Syntax}\label{syntax}
\end{figure}

Statements have access to the local variables and the fields of the enclosing class. 
Statements are standard for sequential composition, assignment, \textbf{if} and \textbf{while} constructs.  
The statement $e.\textbf{get}$, where $e$ is a future variable, blocks the current thread
until $x$ stores the return value.
Evaluation of a right-hand side expression \textbf{new} \emph{C}($\overline{e}$) returns a reference to a new active object within the same group of the executing object,
whereas \textbf{new actor} \emph{C}($\overline{e}$) returns a reference to a new actor 
which forms a new group of active objects.
By $e.m(\overline{e})$ we denote a synchronous method call. Here $e$ is assumed 
to denote an active object, i.e., $e$ is an expression of some type $I$, whereas 
$e!m(\overline{e})$ denotes an asynchronous method call on an actor $e$,
i.e., $e$ is of some type \textbf{Actor}\texttt{<}I\texttt{>}. 

Listing \ref{lst:syn}  contains an example of an actor $bank$ which implements a bank service.
The services provided by a bank are specified by the interface  \texttt{IEmployee}
which is implemented by the class \texttt{Employee}.
A bank is created by a statement

\texttt{Actor <IEmployee> bank = new actor Employee()}.

New employees can be created on the fly by the addEmp method.
The actual data of the bank is represented by the instances of the
class \texttt{Account} which implements the interface \texttt{IAccount} and which contains the  actual methods for transferring credit, checking and withdrawal.
A simple scenario is the following:

\begin{lstlisting}[language=Java, basicstyle=\fontsize{9}{10}\selectfont]
	(1) Fut<Int> f = bank!createAcc(...);
	(2) Int  acc1 = f.get;
	(3) Fut<Bool> f3 = bank!withdraw(acc1, 50);
	(4) Fut<Int> f2 = bank!check(acc1);
\end{lstlisting}

Line 1  models a request to create an account by an asynchronous
method call. The result of this call is a number of the newly created account.
Lines 3 and 4 then describe a withdrawal operation followed by a check on this
account by means of corresponding asynchronous method calls.
These calls are stored in the message queue of the actor $bank$ 
and dispatched for execution by its empoyees, thus allowing a parallel processing
of these requests.
However,  in this particular scenario such a parallel processing of requests
involving the same account clearly may give rise to inconsistent results.
For example a main challenge in this setting arises how to ensure that the messages are \emph{activated}  in the right order, i.e., the order in which they have been queued.
Note that the \emph{execution} of  messages can be synchronized by means of standard synchronization mechanisms, e.g., synchronized methods in Java.
Another approach is to use transactional memory to recover from inconsistent states.
However both approaches do not guarantee in general that the messages are activated in the right order because they do not provide direct control of their {activation}.

By declaring 
in  Listing \ref{lst:syn} all the  parameters of the methods of the interface \texttt{IEmployee} which involve account numbers as synchronized by means of a single
lock ''a''  we ensure  mutual exclusive access to the corresponding accounts.
More specifically,
the selection for execution of a queued message which contains a request  to {withdraw} a certain amount for a specified account,
for example,  requires that (1) no  employee is currently holding the lock ''a'' on that account and (2) no preceding message in the queue requires the lock ''a'' on that account.
Similarly,
a message which contains a transfer request, which involves two accounts, requires that  (1) no  employee is currently holding the lock ''a''  on one of the specified accounts and (2) no preceding message in the queue requires the lock  ''a'' on one of these accounts.
The formal details of this synchronization mechanism is described in the following
section.

\begin{lstlisting}[caption={Syntax Example}\label{lst:syn},language=Java,frame=single,basicstyle=\fontsize{8}{9}\selectfont]
interface IEmployee {
	IAccount createAcc(...);
	Bool transfer(sync<a> Int accNum1, sync<a> Int accNum2, Int amount);
	Bool withdraw(sync<a> Int accNum, Int amount);
	Int check(sync<a> Int acc);
}

interface IAccount {
	Bool transfer(IAccount acc2, Int amount);
	Bool withdraw(Int amount);
	Int check();
}

class Employee implements IEmployee {
	Int createAcc(){
		Int accNum = ...;
		IAccount acc = new Account(accNum, ...); \\account creation
		return accNum;
	}
	Bool transfer(Int accNum1, Int accNum2, Int amount){ 
		IAccount acc1 = getAccount(accNum1);
		IAccount acc2 = getAccount(accNum2);
		acc1.transfer(IAccount acc2, Int amount);...
	}
	Bool withdraw(Int acc, Int amount){
		IAccount acc1 = getAccount(acc1);
		acc.withdraw(Int amount);...
	}
	Int check(Int accNum){
		IAccount acc = getAccount(accNum);
		acc.check();...
	}
	Unit addEmp(){ ...
		IEmployee emp = new Employee();
	}
	IAccount getAccount(Int accNum){...}
}

class Account(Int acn, ...) implements IAccount {
	...
}
\end{lstlisting}

\section{Operational Semantics}\label{opr}
\paragraph{Runtime concepts}
We assume given an infinite set of active object and future references, with typical element $o$
and $f$, respectively. 
We assume distinguished fields \textit{myactor}, \textit{I}, and \textit{L} which denote the identity of the actor, the type of the active object, and the set of pairs of synchronized entries locked by the active object, respectively. 
A local environment $\tau$ assigns values to the local variables (which includes the distinguished variables \textit{this} and \textit{dest},
where the latter is used to store the future reference of the return value). A closure $c=(\tau,s)$ consists of a local environment $\tau$ and a statement $s$.
A thread $t$ is a sequence (i.e., a stack) of closures. A process $p$ of the form $(o, t)$ is a runtime representation of an active object $o$ with an active thread $t$. An actor $a$ denotes a pair $(o,P)$ consisting of an object reference $o$, uniquely identifying the actor as a group of active objects, and a set of processes $P$.  A set $A$ denotes a set of actors. 
By $e$ we denote an event
$m(\bar{v})$ which corresponds to an asynchronous method call with the method name $m$ and values $\overline{v}$. 
For notational convenience, we simply assume that each event also includes information
about the method signature.
A queue $q$ is a sequence of events. 
A (global) context $\gamma$ consists of the following (partial) functions: 
$\gamma_{h}$, which denotes for each existing object its local state, that is,
an assignment of values to its fields; $\gamma_q$, which denotes for  each existing object identifying an actor its queue of events, and, finally,  $\gamma_f$,  which assigns to each existing future  its value ($\perp$, in case it is undefined).

\paragraph{Some auxiliary functions and notations.} 
By  $\gamma[o\leftarrow \sigma]$ we denote  the assignment of the local state
$\sigma$, which assigns values to the fields of $o$,  to the object $o$ (affecting $\gamma_h$);
by $\gamma[o.x\leftarrow v]$ we denote the assignment of  the value $v$ to the field $x$ of object $o$ (affecting $\gamma_h$); by $\gamma[o\leftarrow q]$ we denote the assignment of the queue of events $q$ to the object reference $o$ (affecting $\gamma_q$);
and, finally, by $\gamma[f\leftarrow v]$ we denote the assignment of value $v$ to the future $f$ (affecting $\gamma_f$).
By $\textit{act-dom}(\gamma)$ and $\textit{fut-dom}(\gamma)$ 
we denote the actors and futures specified by the context $\gamma$. 
We assume the evaluation function $val_{\gamma,\tau}(e)$. 
The function $\textit{sync-call}(o, m, \overline{v})$ generates the closure corresponding to a call to the method $m$ of the actor $o$ with the values $\overline{v}$ of the actual parameters.
The function $\textit{async-call}(o,m, \overline{v})$ returns the closure 
corresponding to the message $m(\overline{v})$, where $\bar{v}$ includes
the future generated by the corresponding call (which will be assigned to the local variable \textit{dest}), $o$ denotes the active object
which has been scheduled to execute this method.
In both cases we simply assume that the class name can be extracted from the identity $o$ of the active object (to retrieve the method body). 
The function $\textit{init-act}(o, \overline{v}, o')$ returns the initial state of the new active object $o$. The additional parameter $o'$ denotes the the actor identity which contains $o$, which is used to initialize the field $myactor$ of $o$. The function $\textit{sg}(m(\overline{v}))$ returns the signature of the event $m(\overline{v})$. Finally, $sync_m(\bar{v})$ returns the synchronized arguments of event $m(\bar{v})$ together with their locks (i.e., the arguments specified by \textbf{sync}\texttt{<}$l$\texttt{>} modifier in the syntax where \textit{l} is the lock). 
\paragraph{The Transition Systems}
Figure \ref{local1} gives a system for deriving local transition of the form:
$\gamma,(o, t) \rightarrow \gamma',(o, t')$
which describes the effect of the thread $t$ in the context of $\gamma$. 
Rules \texttt{(ASSIGN-LOCAL)} and \texttt{(ASSIGN-FIELD)} assign the value of expression $e$ to the variable $x$ in the local environment $\tau$ or in the fields $\gamma_h(o')$, respectively. $o'$ is the identity of the active object corresponding to the current closure. Rules \texttt{(COND-TRUE)} and \texttt{(COND-FALSE)} evaluate the boolean expression and branch the execution to the different statements depending on the value from the evaluation of boolean expression $e$. Rule \texttt{(SYNC-CALL)} addresses synchronous method calls between two active objects. A synchronous call gives the control to the callee after binding the values of actual parameters to the formal parameters and forming a closure corresponding to the callee. The closure $(\tau_0, s_0)$, which represents the environment and the statements of the called method, is placed on top of the stack of closures. Rule \texttt{(SYNC-RETURN)} addresses the return from a synchronous method call. We assume that \textbf{return} is always the last statement of a method body. Therefore, the rule consists of obtaining the value $v$ of the return expression $e$, updating the variable which holds the return value on the caller side with $v$, and removing the closure of the callee from the stack.
Rule \texttt{(NEW-ACTOB)} creates a new active object in the same actor by allocating an identity to the new active object and extending the context $\gamma_h$ with the fields of the active object.

Rule \texttt{(READ-FUT)} blocks the active object $o$ until the expression $e$ is resolved, i.e., if $e$ is evaluated to a future which is equal to $\perp$ then the active object blocks.
Rule \texttt{(NEW-ACTOR)} creates a new actor $o'$ and sends the special event $init$ to it with the class name $C$ and the values $\overline{v}$ obtained by evaluating the actual parameters of the constructor. This event will initialize the actor with one active object of type C with the parameters $\overline{v}$. 
Rule \texttt{(ASYNC-CALL)} sends a method invocation message to the actor $o'$ with the new future $f$, the method name $m$, and the values $\overline{v}$ obtained by evaluating the expressions $\overline{e}$ of the actual parameters. The rule updates $\gamma$ to place the message in the queue of the target actor $o'$ and also to extend the set of futures with $f$ with the initial value $\perp$.

Rule \texttt{(SCHED-MSG)} addresses the activation of idle objects of an actor. 
The rule specifies scheduling a thread for the idle object $o$ by binding an event from the queue of the actor $o'$ to which the active object $o$ belongs, and removing the event from the queue. The $q\backslash m(\overline{v})$ removes the first occurrence of message $m(\bar{v})$ from the queue.

The event selection mechanism is underspecified, provided that it respects the temporal order of events in the queue that use the same synchronized data with the same locks. For instance, suppose given events with the order \textit{m1, m2, m3, m4} and \textit{m5} in the queue of an actor with the required set of pairs of lock and data: \textit{\{(l, v1)\}, \{(l', v1)\}, \{(l, v1), (l, v2)\}, \{(l, v2)\},} and \textit{\{(l, v3)\}} for the events respectively (Recall that each synchronized entry is a pair consisting of a data value and a user-defined lock which is specified in the program by the \textbf{sync}\texttt{<}$l$\texttt{>} modifier on the method parameters). The actor also contains more than one active object. If event \textit{m1} is activated then event \textit{m2} can be scheduled in parallel since the required lock for \textit{v1} is different. However, \textit{m3} cannot be scheduled unless \textit{m1} is terminated. Event \textit{m4} also cannot be activated in parallel with \textit{m1}, even though \textit{v2} is free, since \textit{m3} which requires \textit{v2} precedes \textit{m4} in the queue. However \textit{m5} can be activated in parallel with \textit{m1}. The semantics of the \textit{select} function is defined as follows:
\begin{align*}
&\textit{select}(I, L, m(\bar{v}).q) =
\left\{
\begin{array}{ll}
m(\bar{v})& \texttt{in case } L \cap \textit{sync}_m(\bar{v}) = \emptyset \land Sg(m(\bar{v})) \in I\\
\textit{select}(I, L \cup \textit{sync}_m(\bar{v}), q)&
\texttt{otherwise }\\
\end{array}\right.
\end{align*}
where $L \subseteq \textit{Labels} \times \textit{Data}$ and $\textit{select}(I, L, \epsilon) = \perp$ (where $\perp$ stands for undefined).
The signature of selected method requires to be supported by the active object type, $I$. The set of synchronized entries of the message, $\textit{sync}_m(\bar{v})$, also requires to be mutually disjoint with the union of synchronized entries of the actor and the synchronized arguments of the messages preceding to the message in the queue. The binding proceeds then by assigning the set of synchronized entries of the method to the field $L$ of the object. $\textit{Lock}(\gamma, o) = \bigcup \{o'.L | \gamma_h(o'.myactor) = o\}$ returns the synchronized entries of the actor $o$, that is, the union of synchronized entries of its objects, represented by field $L$ of each object. 

Rule \texttt{(ASYNC-RETURN)} evaluates the expression $e$ and assigns the resulting value $v$ to the future $f$ associated to the method call. The return statement belongs to an asynchronous method invocation if there is only one closure in the thread stack (i.e., the closure generated by \texttt{(SCHED-MSG)}). The set $L$ of synchronized entries associated to the invocation are also released by assigning $\emptyset$ to the field $L$ of the active object. Then the closure is removed and the active object $o$ becomes idle. 

\begin{figure}[!ht]
\begin{mathpar}
\\
\inferrule
{\texttt{ASSIGN-LOCAL} \\\\ v = val_{\gamma, \tau}(e) }
{\gamma,(o, t.(\tau,x=e;s))\\\\ \rightarrow \gamma,(o ,t.(\tau[x\leftarrow v],s))}

\inferrule
{\texttt{(ASSIGN-FIELD)} \\\\ o'=\tau(\textit{this}) \\ v = val_{\gamma, \tau}(e)}
{\gamma,(o, t.(\tau,x=e;s))\\\\ \rightarrow \gamma[o'.x \leftarrow v],(o, t.(\tau,s))}

\inferrule
{\texttt{(COND-FALSE)}\\\\ val_{\gamma, \tau}(e) = \text{False}}
{\gamma,(o, t.(\tau,\textbf{if}\ e\ \textbf{then}\ \{s1\}\ \textbf{else}\ \{ s2\} ;s))\\\\ \rightarrow \gamma,(o, t.(\tau,s2;s))}

\inferrule
{\texttt{(SYNC-CALL)}\\\\ o' = val_{\gamma, \tau}(e) \\ \overline{v} = val_{\gamma,\tau}(\overline{e}) \\\\ (\tau_0, s_0) = \textit{sync-call}(o', m, \overline{v})}
{\gamma,(o, t.(\tau,x=e.m(\overline{e});s))\\\\ \rightarrow \gamma,(o, t.(\tau,x=?;s).(\tau_0, s_0))}

\inferrule
{\texttt{(COND-TRUE)}\\\\ val_{\gamma, \tau}(e) = \text{True}}
{\gamma,(o, t.(\tau,\textbf{if}\ e\ \textbf{then}\ \{s1\}\ \textbf{else}\ \{ s2\} ;s))\\\\ \rightarrow \gamma,(o, t.(\tau,s1;s))}

\inferrule
{\texttt{(SYNC-RETURN)}\\\\ v = val_{\gamma, \tau}(e)}
{\gamma,(o, t.(\tau,x=?;s).(\tau_0, \textbf{return}\ e)) \rightarrow \gamma,(o, t.(\tau,x=v;s))}

\inferrule
{\texttt{(READ-FUT)} \\\\val_{\gamma, \tau}(e) \neq \perp}
{\gamma,(o, t.(\tau, e.\textbf{get};s)) \rightarrow \gamma, (o, t.(\tau, s))}

\inferrule
{\texttt{(NEW-ACTOB)}\\\\ o' \not\in \textit{dom}(\gamma_h)}
{\gamma,(o, t.(\tau ,x=\textbf{new }C(\overline{e});s)) \rightarrow \gamma[o'\leftarrow \textit{init-act}(o', val_{(\gamma,\tau)}(\overline{e}), \gamma_{h}(o.\textit{myactor}))] ,(o, t.(\tau[x\leftarrow o'] ,s))}

\inferrule
{\texttt{(NEW-ACTOR)}\\\\ o' \not\in \textit{act-dom}(\gamma) \\ \overline{v} = val_{\gamma,\tau}(\overline{e})}
{\gamma, (o, t.(\tau ,x=\textbf{new actor }C(\overline{e});s)) \rightarrow \\\gamma[o'\leftarrow \textit{init}(C, \overline{v})],(o, t.(\tau[x\leftarrow o'] ,s))}

\inferrule
{\texttt{(ASYNC-CALL)}\\\\ f \notin \textit{fut-dom}(\gamma) \\ \overline{v} = val_{\gamma,\tau}{(\overline{e})} \\ o' = val_{\gamma,\tau}{(e)} \\ \gamma_q(o')= q}
{\gamma, (o, t.(\tau,x=e!m(\overline{e});s)) \rightarrow \gamma[f\leftarrow \perp, o'\leftarrow q.m(\overline{v}, f)],(o, t.(\tau,x=f;s))}

\inferrule
{\texttt{(ASYNC-RETURN)}\\\\ v = val_{\gamma, \tau}(e) \\ f= \tau(\textit{dest})}
{\gamma,(o, (\tau, \textbf{return}\ e)) \rightarrow \gamma[f\leftarrow v, o.L \leftarrow \emptyset], (o, \epsilon)}

\inferrule
{\texttt{(SCHED-MSG)}\\\\ o' = \gamma_h(o.\textit{myactor}) \\ \gamma_q(o') = q \\ m(\overline{v}) = select(\gamma_h(o.\textit{I}), lock(\gamma, o'), q) \\ (\tau, s) = \textit{async-call}(o, m, \overline{v})}
{\gamma,(o, \epsilon) \rightarrow \gamma[o' \leftarrow q\backslash m(\overline{v}), o.L \leftarrow \textit{sync}_m(\overline{v})], (o, (\tau, s))}
\end{mathpar}
\caption{Operational Semantics at the Local Level}
\label{local1}
\end{figure}
Figure \ref{osgroup} gives the rules for the second level, the actor level. Rule \texttt{(PROCESS-} \texttt{UPDATE)} specifies that if the domain of the heap remains the same then only the current process is updated. Rule \texttt{(PROCESS-CREATE)}, on the other hand, shows that if the domain of the heap has been extended with a new active object $o'$ then a new idle process $p''$ for the active object $o'$ is introduced to the processes of the actor.

Figure \ref{ossystem} gives the rules for the third level, the system level. Rule \texttt{(ACTOR-} \texttt{UPDATE)} specifies that if the domain of $\gamma$ remains the same then only the current actor is updated. Rule \texttt{(ACTOR-CREATE)}, on the other hand, shows that if the domain of $\gamma$ has been extended 
then a corresponding new actor configuration $a''$ is added to the system. Note that this actor is identified by the reference which has been added to $\gamma$.
This reference is also used to identify the initial active object of the newly created actor.
\begin{figure}[!htbp]
\begin{mathpar}
\inferrule
{\texttt{(PROCESS-UPDATE)}\\\\\gamma, p \rightarrow \gamma', p' \\\\ \textit{dom}(\gamma_h) = \textit{dom}(\gamma_{h'})}{\gamma, (o, P\cup \{p\})\rightarrow \gamma', (o, P\cup \{p'\})}

\inferrule
{\texttt{(PROCESS-CREATE)}\\\\\gamma, p \rightarrow \gamma', p' \\\\ o' \in \textit{dom}(\gamma_{h'})\backslash \textit{dom}(\gamma_h) \\ p'' = (o', \epsilon)}{\gamma, (o, P\cup \{p\})\rightarrow \gamma', (o, P\cup \{p', p''\})}
\end{mathpar}
\caption{Operational Semantics at the Actor Level}
\label{osgroup}
\end{figure}
\begin{figure}
\begin{mathpar}
\inferrule
{\texttt{(ACTOR-UPDATE)}\\\\\gamma, a\rightarrow \gamma', a' \\\\ \textit{act-dom}(\gamma) = \textit{act-dom}(\gamma')}
{\gamma, A\cup \{a\}\rightarrow \gamma', A\cup \{a'\}}

\inferrule
{\texttt{(ACTOR-CREATE)}\\\\\gamma, a\rightarrow \gamma', a' \\ o \in \textit{act-dom}(\gamma')\backslash \textit{act-dom}(\gamma) \\\\ \gamma'_q(o) = q.\textit{init}(C, \overline{v}) \\ a'' = (o, \{(o, \epsilon)\})}
{\gamma, A\cup \{a\}\rightarrow \\\\ \gamma'[o\leftarrow q, o\leftarrow \textit{init-act}(o, \overline{v}, o)], A\cup \{a', a''\}}
\end{mathpar}
\caption{Operational Semantics at the System Level}
\label{ossystem}
\end{figure}

We have the following the description of the initial state for the operational semantics in the local, actor, and system level respectively:
\begin{equation*}
p_0 = (\_ , (\tau_{main}, s_{main})) \qquad a_0 = (\_, \{p\})\qquad A_0 = \{a\}
\end{equation*}
The $p_0$ represents a process with the context $\tau_{main}$ for the main body and its statement $s_{main}$. The process is considered to be an active object with the anonymous identity which is denoted by underscore. 
The $a_0$ represents an anonymous actor with the underscore identity in the system and the process $p_0$ in the process set. The gamma is initialized as the following, $$ \gamma[\_ \leftarrow \{myactor \leftarrow \_\}]$$ as the active object state for $p_0$. Any object which is created in the main body is a free object, an active object that belongs to the anonymous actor. All the objects which are created by a free object are also free objects. The field \textit{myactor} of all the free objects is equal to underscore. The anonymous actor does not receive any event as it has no identity in the program. 

We conclude this section with the following basic operational property of synchronized data:
\begin{theorem}
First, let $\textit{Object}(a) = \{ o\mid (o, t) \in P,\; \mbox{\rm for some process $p$} \}$ denote the set of objects in $a$ which contains the set processes $P$. 
For every configuration $A$ reachable from the initial configuration $A_0$ we have $o.L \cap o'.L = \emptyset$ for any $o, o' \in \textit{Object}(a)$ $(o\neq o')$ 
\end{theorem}

This invariant property follows immediately from the definition of the select function. It expresses that at run-time there are no two distinct asynchronous method invocations which require the same synchronized data.

\section{Experimental Methodology and Implementation}
\label{exp}
In this section we present the implementation of the MAC in a widely used, mainstream programming language, the Java language. The implementation has to take into account the transparency of parallel computation from the user's perspective and the functions that are exposed by the abstract class.
The outline of the implementation is presented in Listing \ref{lst:ogc}.

As shown in the operational semantics in section \ref{opr}, the default policy schedules the idle objects non-deterministically. However, there is the possibility to overload the policy using the runtime information to allow a preferential selection of the active objects. Furthermore, the current selection method presented is minimal, in the sense that it can be overloaded with different arguments to provide more selection options based on application specific requirements. 

\subsection{Actor Abstract Class}

\par The Java module creates an abstract class, \textbf{Actor}, that provides a runtime system for queuing and activation of messages. It exposes two methods to the outside world for interaction, namely \textit{send(Object message)} and \textit{getNewWorker(Object... parameters)}. This layout is used to allow a clear separation between internal object selection, message delivery and execution. This class is the mediator between the outside applications and the active objects defined by the internal interface \textbf{ActiveObject}. These Active Objects will be assigned execution of the requests sent to the actor. Our abstract class contains a queue of \textit{availableWorkers} and a set of \textit{busyWorkers} separating those objects that are idle from those that have been assigned a request. Parallel execution and control is ensured through a specific Java Fork Join Pool \textit{mainExecutor} that handles the invocations assigned to the internal objects and is optimized for small tasks executing in parallel. The class uses a special queue, named \textit{messageQueue}, that is independent of the thread execution. It is used to store incoming messages and model the \textbf{shared queue} of the group. This message queue is initialized with a comparator(ordering function) that selects the first available message according to the rule \texttt{(SCHED-MSG)} specified in Section \ref{opr}. To use this abstract class as a specific model, it needs to be extended by each interface defined in our language in order to be initialized as an Actor.

\par The default behavior of the exposed method \textit{getNewWorker(Object... parameters)} is to select a worker from the \textit{availableWorkers} queue. The workers are inserted in a first-in-first-out(FIFO) order with a blocking message delivery if there is no available worker (i.e. the \textit{availableWorkers} queue is empty). While the behavior of this method is hidden from the user, it needs to be exposed such that the user has a clear view of the selection, before sending a request. 
The second exposed function, \textit{send(Object message, Set\texttt{<}Object\texttt{>} data}, takes the first argument in the form of a lambda expression, that models the request. The format of the lambda expression must be \begin{verbatim}
() -> ( getNewWorker() ).m()
\end{verbatim}
The second argument specifies a set of objects that the method \textit{m()} needs to lock and maintain data consistency on. Therefore when a request is made for a method \textit{m()} the runtime system must also select an \textit{Active Object} from the \textit{availableWorkers} queue to be assigned the request, as well a set of data that needs concurrency control. Execution is then forwarded to the \textit{mainExecutor} which returns a Java Future to the user for synchronization with the rest of the application outside the actor. The selection of the \textit{ActiveObject} is important to form the lambda expression that saves the application from having a significant number of suspended threads if the set of data that is required is locked.

\par The application outside of the Actor sends requests asynchronously and must be free to continue execution regardless of the completion of the request. To this end we provide the class \textbf{Message} illustrated in Listing \ref{lst:mess} which creates an object from the arguments of the \textit{send} method and initializes a future from the lambda expression. 

\begin{lstlisting}[float=!ht,caption={Message Class in Java}\label{lst:mess},language=Java,basicstyle=\fontsize{7}{9}\selectfont, frame=single, tabsize=4]
import java.util.Set;
import java.util.concurrent.Callable;
import java.util.concurrent.ForkJoinTask;
import java.util.concurrent.atomic.AtomicInteger;

public class Message {

	static final AtomicInteger queuePriority = new AtomicInteger(0);

	String name;
	Object lambdaExpression;
	Set<Object> syncVariables;
	ForkJoinTask<?> f;
	AtomicInteger preemptPriority;
	int priority = 0;

	public Message(Object message, Set<Object> variables, String name) {
		this.lambdaExpression = message;
		this.syncVariables = variables;
		this.name = name;
		this.preemptPriority = new AtomicInteger(0);
		priority = queuePriority.getAndAdd(1);

		f = null;
		if (message instanceof Runnable)
			f = ForkJoinTask.adapt((Runnable) message);
		if (message instanceof Callable<?>)
			f = ForkJoinTask.adapt((Callable<?>) message);
	}

	public Message(Message m) {
		this(m.lambdaExpression, m.syncVariables, m.name);
		this.preemptPriority.set(m.preemptPriority.get());
	}

	@Override
	public String toString() {
		return name + " " + syncVariables + " :< " + priority + ","
				+ preemptPriority + " >";
	}
}
\end{lstlisting}

This class contains the specific parts of a message which are the \textit{lambdaExpression}, the \textit{syncData} on which the request need exclusive access and the Future \textit{f} which captures the result of the request. To maintain a temporal order on messages synchronize on the same messages the class also contains a static field \textit{queuePriority} which determines a new message's \textit{priority} upon creation and insertion in the queue.

\par The Actor runs as a process that receives requests and runs them in parallel while maintaining data-consistency throughout its lifetime. The abstraction is data-oriented as it is a stateful object maintaining records of all the data that its workers are processing. It contains a set of \textit{busyData} specifying which objects are currently locked by the running active objects. An internal method, named \textit{reportSynchronizedData} is defined to determine if a set of data corresponding to a possible candidate message for execution is intersecting with the current set of \textit{busyData}. This method is used as part of the comparator defined in the \textit{messageQueue} to order the messages based on their availability. The main process running the Actor is then responsible to take the message at the head of the queue and schedule it for execution and add the data locked by the message to the set of \textit{busyData}. It is possible that at some point during execution, all messages present in the \textit{messagesQueue} are not able to execute due to their data being locked by the requests that are currently executing. To ensure that our Actor does not busy-wait, we forward all the messages into a \textit{lockedQueue} such that the Actor thread suspends.

\par The Actor is a solution that makes parallel computation transparent to the user through the internal class implementation of its worker actors. These objects are synchronized and can undertake one assignment at a time. Each request may have a set of synchronized variable to which it has exclusive access while executing. At the end of the execution, the active object calls the \textit{freeWorker(ActiveObject worker, Object ... data)} method that removes itself from the \textit{busyWorkers} set and becomes available again by inserting itself in the \textit{availableWorkers} queue. At this point, the \textit{lockedQueue} is flushed into the \textit{messageQueue} such that all previously locked messages may be checked as candidates for running again. All of the objects that were locked by this ActiveObject are also passed to this method such that they can be removed from the \textit{busyData} set and possibly release existing messages in the newly filled \textit{messageQueue} for execution. This control flow is illustrated in an example in the next section, however our motivation is to modify this module into an API and use it as a basis for a compiler from the modeling language to Java.

\subsection{Service Example and Analysis}

\par Listing \ref{lst:bank} shows the implementation of a \textbf{Bank} service as an \textbf{Actor}. 
As a default behavior, whenever a new concrete extension of an Actor is made, the constructor or the \textit{addWorkers} method may create one or more instances of the internal Active Object.
\begin{lstlisting}[float=h,caption={Bank Class in Java}\label{lst:bank},language=Java,basicstyle=\fontsize{7}{7}\selectfont, frame=single, tabsize=4]
public class Bank extends Actor {

	public void addWorkers(int n){
		for (int i = 0; i < n; i++) {
			availableWorkers.add(new BankEmployee());
		}
	}
	
	@Override
	public BankEmployee getNewWorker(Object... parameters) {
		BankEmployee selected_worker = null;
		try {
			selected_worker = (BankEmployee) availableWorkers.take();
			busyWorkers.add(selected_worker);
		} catch (InterruptedException e) {
			e.printStackTrace();
		}
		return selected_worker;
	}

	class BankEmployee implements ActiveObject {

		public Account createAccount() {
			Account a = new Account();
			Bank.this.freeWorker(this);
			return a;
		}

		public boolean withdraw(Account n, int x) {
			boolean b = n.withdraw(x);
			Bank.this.freeWorker(this, n);
			return b;
		}

		protected boolean deposit(Account n, int x) {
			boolean b = n.deposit(x);
			Bank.this.freeWorker(this, n);
			return b;
		}

		public boolean transfer(Account n1, Account n2, int amount) {
			boolean b = n1.transfer(n2, amount);
			Bank.this.freeWorker(this, n1, n2);
			return b;
		}

		public int checkSavings(Account n) {
			int res = n.checkSavings();
			Bank.this.freeWorker(this, n);
			return res;
		}

		class Account {
			//Acount processing methods
}	}	} 	}
\end{lstlisting}
The behavior of \textit{getNewWorker(Object... parameters)} is overridden to ensure the return of a specific internal Active Object with exposed methods, in this case the \textbf{BankEmployee}. This internal class implements the general \textit{Active Object} interface and exposes a few simple methods of a general Bank Service. The methods \textit{withdraw, deposit, transfer} and \textit{checkSavings} all perform their respective operations on one or more references of the internal class Account a reference which is made available through the method \textit{createAccount}.  The MAC behavior is inherited from the Actor and only the specific banking operations are implemented.

\par
\begin{figure}[!ht]
\centering
\begin{minipage}{.99\textwidth}
	\centering
\includegraphics[scale=0.4]{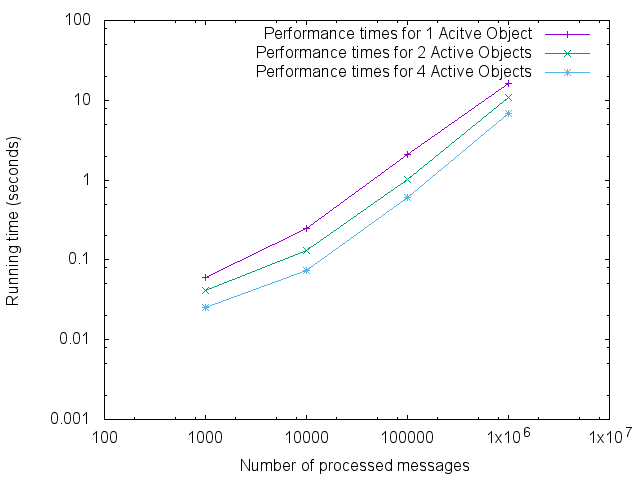}
\caption{Performance Times for processing 100-1M messages}
\label{fig:time}
\end{minipage}
\begin{minipage}{.99\textwidth}
	\centering
	\includegraphics[scale=0.4]{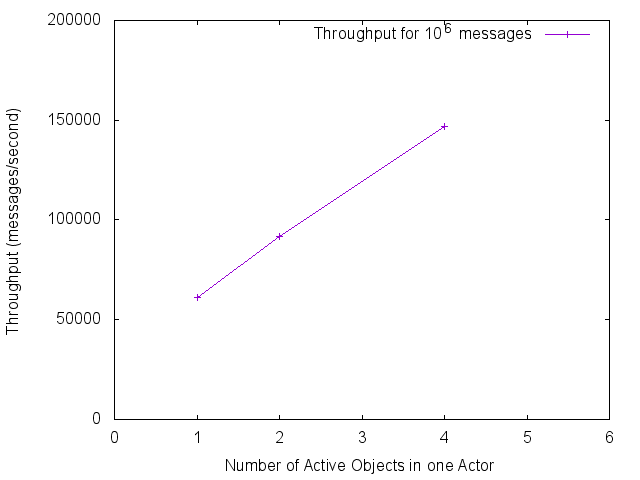}
	\caption{Throughput for processing 1M messages}
	\label{fig:th}
\end{minipage}
\end{figure}

 To test the functionality, as well as the performance of the MAC we implement a simple scenario that creates a fixed number of users each operating on their own bank account. We issue between 100 and 1 million requests distributed evenly over the fixed number of accounts. To ensure that some messages have to respect a temporal order and forced await execution of prior requests on the same account we issue sets of 10 calls for each account. This also ensures that the selection rule \texttt{(SCHED-MSG)} does not become too large of a bottleneck as in the case of issuing all operations for one bank account at a time.  We measure the time taken to process the requests based on a varying number of Active Objects inside in the \textbf{Bank} Service. The performance figures for a MAC with 1,2 and 4 available \textit{Active Objects} is presented in Figure \ref{fig:time}

The results validate our solution in the sense that the time:message ratio is almost linear with very little overhead introduced by the message format and the selection function. Furthermore the benefit of parallelism is maintained with the increasing volume of request issued to the service. To emphasize this we computed the throughput of the service in relation to the number of \textit{Active Objects} running and present it in Figure \ref{fig:th}. From these results we can infer the scalability of the MAC for parallel computation.

\begin{lstlisting}[float=!ht,caption={Actor Abstract Class in Java}\label{lst:ogc},language=Java,basicstyle=\fontsize{7}{8}\selectfont, frame=single, tabsize=4]
public abstract class Actor implements Runnable {

	protected ForkJoinPool mainExecutor;
	protected PriorityBlockingQueue<Message> messageQueue;
	protected ConcurrentLinkedQueue<Message> lockedQueue;
	protected PriorityBlockingQueue<ActiveObject> availableWorkers;
	protected Set<ActiveObject> busyWorkers;
	protected Set<Object> busyData;

	public Actor() {
		//initialization of internal data structures
	}

	@Override
	public void run() {
		
		while (true) {
			try {
				Message message = messageQueue.take();
				if (reportSynchronizedData(message.syncData)) {
					synchronized (busyData) {
						busyData.addAll(message.syncData);
					}
					mainExecutor.submit(message.f);
				} else {
					this.lockedQueue.offer(newM);
				}
			} catch (InterruptedException e) {
				e.printStackTrace();
			}
	}}

	// message format: ()->getWorker().m()
	public <V> Future<V> send(Object message, Set<Object> data) {
		Message m = new Message(message, data, name);
		messageQueue.put(m);
		return (Future<V>) m.f;
	}

	public ActiveObject getNewWorker(Object... parameters) {
		ActiveObject selected_worker = null;
		
		try {
			selected_worker = availableWorkers.take();
			busyWorkers.add(selected_worker);
		} catch (InterruptedException e) {
			e.printStackTrace();
		}

		return selected_worker;
	}

	private boolean reportSynchronizedData(Set<Object> data) {
		Set<Object> tempSet = new HashSet<Object>();

		synchronized (busyData) {
			tempSet.addAll(busyData);
			tempSet.retainAll(data);

			if (tempSet.isEmpty()) {
				return true;
			}
			return false;
	}	}

	protected void freeWorker(ActiveObject worker, Object... data) {

		synchronized (busyData) {
			busyWorkers.remove(worker);
			availableWorkers.offer(worker);
			
			messageQueue.addAll(lockedQueue);
			lockedQueue.clear();
			
			
			for (Object object : data) {
				busyData.remove(object);
	}	}	}

	public interface ActiveObject extends Comparable<ActiveObject> {
		// the active objects in charge of requests
	}
}
\end{lstlisting}

\section{Conclusion and Future Work}
\label{concl}
In this paper we have introduced the notion of multi-threaded actors, that is, an
actor-based  programming abstraction which allows to model an actor as  a group of  active objects which share a message queue.   
Encapsulation of the active objects which share a queue can be obtained by simply
not allowing active objects to be passed around in asynchronous messages.
Cooperative scheduling of the method
invocations within an active object  (as described in for example \cite{JohnsenO07}),
can be obtained by introduction of a lock for each active object.
In general,  synchronization mechanisms between threads  is an orthogonal issue
and as such can be easily integrated, e.g., lock on objects, synchronized methods (with reentrance), or even synchronization by the compatibility relationship between methods as defined in \cite{haustein2006jac} and \cite{henrio2013multi}. Other extensions and variations describing dynamic
group interfaces can be considered along the lines of \cite{Johnsen2016}.

Future work will be dedicated toward the development of the compiler which allows importing Java libraries, and further development of the runtime system, as well as benchmarking on the performance. 
Other work of interest is to investigate into dynamic interfaces for the multi-threaded actors  and programming abstractions for  application-specific scheduling
of multi-threaded actors.
\nocite{*}
\bibliographystyle{eptcs}
\bibliography{mac}
\end{document}